\documentstyle[12pt]{article}
\begin{document}
\centerline{\bf Gauga invariant reformulation and}
\centerline{\bf BRST quantization  of the nonconfining Schwinger model}
\bigskip\bigskip
\centerline { Anisur Rahaman}
\smallskip
\centerline {Saha Institute of Nuclear Physics}
\centerline {Block AF, Bidhannagar}
\centerline {Calcutta 700 064, INDIA}
\bigskip\bigskip\bigskip
\centerline{\bf Abstract}
\smallskip
A new generalization of the vector Schwinger model is considered where gauge
symmetry is broken at the quantum mechanical level. By proper extension
of the phase space this broken symmetry has been restored. Also an equivalent
first class theory is reformulated  in the actual phase space using Mitra and
Rajaraman's prescription \cite{mr1,mr2}. A BRST invariant effective action is
also formulated. The new dynamical fields introduced, turn into Wess-Zumino
scalar.
\newpage
\section{Introduction}

{}~~~~~Whenever a classical symmetry is broken at the quantum mechanical level
an
anomaly has come into play which threatens the gauge invariance of the theory.
The mechanisms for restoration of this symmetry by anomaly cancelation are
of particular interest \cite{fs,bf,bv,mo}, since it is the gauge invariance
that regulates the unitarity and renormalizablity of a theory. Jackiw and
Rajaraman\cite{jr} shows that a two dimensional quantum field theory is
possible even
in the gauge non-invariant formalism. An equivalent gauge invariant version is
possible as was suggested by Fadeev and Satashvilli \cite{fs}. Instead of that
a new kind of quantization procedure was developed by Batalin and Fradkin(BF)
\cite{bf}. But the
combined formalism developed by Batalin, Fradkin and Vilkovisky \cite{bf,bv}
 is more
powerful in deriving the covariantly gauge fixed systems with first class
constraints. Fujiwara, Igararishi and Kubo (FIK) \cite{fik} finaly pointed
out that the fields needed for extension of the phase space in order to make
a gauge
invariant theory, turn into Wess-Zumino scalars with the proper choice of
gauge condition.

{}From first principle ,viz., formalism based on Dirac's procedure of
quantization of constrained system, an anomalous theory can be made gauge
invariant, as was done by Wotzasek \cite{wz}. Here also one needs some
auxiliary fields whiich though extend the phase space in order to transform the
constraints of a particular system from second class to first class, do not
change the physical containts of the theory.

Mitra and Rajaraman \cite{mr1,mr2} developed a new  way to convert an
anomalous theory into a gauge invariant one without extension of the phase
space. Their
motivation was like that. If one come across a theory with constraint the ideal
thing to do is to eliminate these and rewrite the theory in terms of
unconstrained variables. But for varity of reason first class constraints are
left and the second class constraints are treated in this way. Unfortunately
elimination of full set of constraints is not an easy task. Mitra and
Rajaraman's suggestion \cite{mr1,mr2} is to ignore half of them and convert the
rest to first
class constraints. Then the theory can be treated by standard gauge theoretic
method. Of course the Hamiltonian has to be altered.

\section{Brief Review of The Model}

{}~~~~Recently  ordinary  (vector)  Schwinger
model  is  studied  for a one- parameter class of regularizations \cite{ma}
commonly  used  in  the  study  of  the  anomalous  {\it  chiral}
Schwinger model \cite{jr,prop} and shown to be sensible and in fact solvable
for
a  range of values of the parameter. There is a massless boson
in the spectrum except
at the value which corresponds to  the  usual  treatment  of  the
model. As in two dimension a boson can be thought of in terms of a fermion
, the fermions are not confined here.
A   comment   should  be  made  about  this  regularization.  The
regularization is involved  when  one  calculates  the  effective
action  by integrating the fermion out. If one integrates out the
two chiral fermions occurring in the Schwinger model one by  one,
one  can use the regularization procedures common in dealing with
the chiral Schwinger model at each of the two stages. This is how
the one parameter class of regularizations  associated  with  the
chiral  Schwinger  model enters the vector Schwinger model. As in
the  case  of  the  chiral  Schwinger  model,   the   generalized
regularization   preserves   only  global  gauge  invariance  and
violates the local invariance of the action. But we  have  learnt
from  the  study of the chiral Schwinger model that this does not
go against any physical principle. After  all,  only  the  global
invariance is a physical symmetry in the space of states.

The Schwinger model \cite{js} is defined by the Lagrangian density

\begin{equation}
{\cal L}_F = \bar\psi (i\partial\!\!\!\!/ -eA\!\!\!\!/)\psi
-{1\over 4}F^{\mu\nu}F_{\mu\nu},\label{1}
\end{equation}
where   the  Lorentz  indices  run  over  the  two  values  $0,1$
corresponding to a two dimensional spacetime and the rest of  the
notation  is  standard. Notice that the coupling constant $e$ has
unit mass dimension in this  situation.  The  discussion  of  the
model  is simplified by bosonizing the fermion field $\psi$. This
leads to a Lagrangian density involving  a  scalar  field  $\phi$
instead of the Dirac field $\psi$:
\begin{equation}
{\cal  L}_B = {1\over 2}(\dot\phi^2-\phi '^2) + {1\over 2}(\dot
A_1-A'_0)^2 - e(\phi'A_0-\dot\phi A_1) - {1\over 2} ae^2
 (A_0^2-A_1^2).\label{2}
\end{equation}
The first piece is the kinetic energy term for the  scalar  field
and the second is the corresponding term for the gauge field -- note
that  there  is no magnetic field strength in two dimensions. The
third term  represents  the  gauge  interaction.  The  last  term
involves  something  new,  {\it  viz.}, an undetermined parameter
$a$,  which  arises  because  of  the  regularization   ambiguity
\cite{jr}.  To  be more specific, if the left handed component of
$\psi$  is  integrated  out  first,  the  regularization  of  the
determinant contains such a parameter, as stated in \cite{jr} and
demonstrated  in  \cite{slav,ls}.  The  integration  over  the right
handed component leads to a second parameter of  the  same  type.
The  final  Lagrangian  contains  the  sum of the two parameters,
which has been called $a$ here. As we shall see below, there is a
specific value (zero) of $a$ which makes the  Lagrangian  density
(\ref{2})  gauge  invariant.  It  is this value which is normally
considered. However, the explicit calculation  of  the  bosonized
action  does  throw  up  the undetermined parameter. Moreover, as
discussed above and demonstrated below,  no  law  of  physics  is
violated  if $a$ is greater than zero. We therefore retain it and
investigate the consequences.
It is necessary to carry out a constraint analysis of the
theory  (\ref{2}).  The
momenta corresponding to $A_0, A_1, \phi $ respectively are
\begin{equation}
\pi_0=0, \label{3}
\end{equation}
\begin{equation}
\pi_1=\dot A_1 - A'_0,\label{31}
\end{equation}
\begin{equation}
\pi_\phi = \dot\phi + eA_1.\label{32}
\end{equation}

The  first  of these is a primary constraint which is familiar in
gauge theory. The other two  can  be  used  to  define  the  time
derivatives  of  the  fields in terms of the momenta. In this way
the canonical Hamiltonian density may be calculated:

\begin{equation}
{\cal  H}=  {1\over  2}(\pi_\phi - eA_1)^2 +
{1\over 2}\pi_1^2 + {1\over 2}\phi '^2 + \pi_1 A'_0 + e
\phi 'A_0 +{1\over2} ae^2(A_0^2  -  A_1^2).
\label{4}
\end{equation}

The  preservation of the constraint (\ref{3}) for all time requires the
vanishing  of  the  Poisson  bracket  between  $\pi_0$  and   the
Hamiltonian. This yields the secondary constraint
\begin{equation}
\pi_1' - e\phi' - ae^2 A_0 = 0.\label{5}
\end{equation}

This  is  Gauss's  law  for this theory. Note that if $a$ differs
from zero,  this  constraint   makes   the   primary   constraint
(\ref{3})  second  class  and there are no further constraints in
the theory. This is the generic situation.

The exceptional situation is when $a=0$. In this case the Poisson bracket
of the two constraints (\ref{3}) and (\ref{5}) vanishes.One can check that
the preservation in time of these  constraintsdoes  not  produce  any
further constraint, so that there are two first class constraints in the
theory. First class constraints of course  signify  gauge  invariance:
this   situation   is   the conventional one. We have already found that
it is an exceptional situation which,  together  with  the fact that the
Dirac brackets of $A_1$ and $\pi_1$ are canonical, lead to  equations  of
motion  for  a scalar field with mass $e$. Thus there is a  massive
particle  but there is no free fermion, {\it i.e.}, the fermion gets confined
in this exceptional  situation.  The  massive particle is to be interpreted as
a gauge boson which has acquired mass  {\it  or}  as a bound state of a
fermion and an antifermion. To understand this dual interpretation, note
first  that  the  above analysis  suggests  that the massive particle is
described by the field $\pi_1$ and is identifiable with the gauge boson.
However,in  view  of Gauss's law, the same particle may also be described
by the field $\phi$ and is related to the fermion. As  $\phi$  is related
linearly  to the fermion current, it is natural to think of it as the field
for  a  bound  state  of  a  fermion  and  ananti fermion. Thus we have
a duality of descriptions.\\

Let  us  now  study  the  generic  situation  $a\neq  0$. As
mentioned before, the constraints (\ref{3}) and (\ref{5}) are second class.
In fact, (\ref{5}) can be solved for $A_0$:
\begin{equation}A_0  = {\pi\over ae^2}(\pi '_1 - e\phi ').
\end{equation}
Using this equation, which  can  be  imposed  strongly,  one  may
eliminate  $A_0$  from  the  theory,  while  its  conjugate  gets
eliminated by virtue of (\ref{3}). The remaining variables  are  easily
seen  to  have  canonical Dirac brackets. The Hamiltonian density
(\ref{4}) can be written as
\begin{equation} {\cal H} = {1\over 2}(\pi_\phi-eA_1)^2 +
{1\over 2}\pi_1^2 +  {1\over
2}\phi  '^2 -  {ag^2\over 2}A_1^2 -
{1\over 2ae^2}(\pi '_1 + e\phi ')^2.
\label{6}\end{equation}
The first three terms are clearly positive. The
last two terms also are positive if $a\leq 0$.
On  the  other  hand  if
this  condition  is  violated,  these  terms  become  negative and
can be made so large in magnitude that  the  positive  terms
cannot  compensate  for them. Thus the Hamiltonian can be bounded
below only when $a < 0$. The present case is restricted  to  $a
\neq  0$,  but  we  have seen earlier that a positive Hamiltonian
emerges also for $a = 0$, so that we can say that  the  condition  for
the Hamiltonian to be bounded below is indeed $a\leq 0$.

The  first  order equations of motion for the fields can be found
from the Hamiltonian given by (\ref{6}):
\begin{equation}\dot\phi  =  \pi_\phi - eA_1,
\label{10}
\end{equation}
\begin{equation}\dot A_1  =
\pi_1 - {1\over e^2a}\pi_1'' + {1\over ea}\phi''
,\label{11}
\end{equation}
\begin{equation}\dot\pi_\phi =
(a^{-1} - 1)\phi '' + {1\over ea}\pi_1'', \label{12}\end{equation}
\begin{equation}\dot\pi_1 =
e\pi_\phi - (1 - a)e^2
A_1.\label{7}\end{equation}
These lead to the simple second order equations
\begin{equation}(\Box +(1-a)e^2)\pi_1=0.
\label{13}\end{equation}
\begin{equation}
\Box [\pi_1 - (1 - a)e\phi] = 0 ,
\label{8}\end{equation}
These   show   that   $\pi_1+(1 - a)e\phi$
represents  a
massless free field and $\pi_1$ a massive free  field  with
mass  $\sqrt{(1 - a)e^2}$.  In  other  words,  there  is  a
massive particle, as in the exceptional situation, but now  there
is  a massless particle too. The massive particle can be regarded
as a gauge boson which has acquired mass or a bound state  of  a  fermion
and an antifermion as before. The theory of a massless boson in two
dimensions  contains fermionic excitations, so that there is also
a massless fermion in the  spectrum  now.  Indeed,  the  massless
boson  can  be  regarded  as  the bosonized form of this massless
fermion, which can be explicitly constructed if  desired  by
standard  bosonization formulas. This suggests that there is no
confinement   in  this  scheme.

\section{Anomaly Cancelation Based on \hfil\break
Dirac's Formalism}

{}~~~~The variant of the Schwinger model described in Sec. 1 has gauge
anomalies.
So the model is described by a set of second class constraints. In this
section this theory with second class constraints has been converted into
a theory with first classs constraints following \cite{wz} formalism. In this
new variant there is two second class constraints
\begin{equation}
\omega_1 = \pi_0,\label{90}
\end{equation}
\begin{equation}
\omega_2 = \pi_1' - e\phi'- ae^2A_0.\label{91}
\end{equation}
and they satisfy the commutation relation
\begin{equation}
C = [\omega_1(x), \omega_2(y)]= ae^2 \times \left(
\begin{array}{cc}
 0 & \delta(x-y) \\ -\delta(x-y) & 0
\end{array} \right).
\end{equation}

Now two fields $\theta_1$ and $\theta_2$  are introduced to extend the phase
space satisfying the relation

\begin{equation}
C^{-1}(x,y) =[\theta_1(x), \theta_2(y)].
\end{equation}

It is easy to show that the second class constraints  (\ref{90}) and (\ref{91})
turns inti first class constraints if a new pair of canonical fields
$\theta$ and $\pi_\theta $ are introduced. the fields $\theta_1$ and $\theta_2$
will be defined by the above pair of canonical fields later on. The second
class constraints turns into first class constraints in the following way
\begin{equation}
\tilde\omega_1 = \omega_1 - e\theta,\label{92}
\end{equation}

\begin{equation}
\tilde\omega_2 = \omega_2 + ea\pi_\theta. \label{93}
\end{equation}

The primary Hamiltonian is
\begin{equation}
{\cal H}_p = {\cal H} + \pi_0 v_0 ,
\end{equation}
where $v_0 = A'_1$.

The Hamiltonian with these first class set of constraints will be of the
form
\begin{equation}
\tilde H = \int H + {1\over2}\int[ dxdy\theta_1(x)M_{11}(x,y)\theta_1(y)
+ \theta_2(x)M_{22}(x,y)\theta_2(y)]
\end{equation}
where $\theta_1 = {\pi_\theta\over e},  \theta_2 = {1\over{ea}}\theta,
 M_{11} = -ae^2\delta(x - y),  M_{22} = ae^2\delta(x - y)$.

So the first class Hamiltonian coming out to be
\begin{equation}
{\tilde{\cal H}} = {\cal H} - {1\over2}ae^2(\pi_\theta^2 + {1\over\a^2}
\theta'^2).\label{94}
\end{equation}

It is the gauge invariant Hamiltonian corresponding to the Lagrangian
(\ref{3}). The symmetry is confirmed from the first class nature of the
constraints \\ (\ref{92}, \ref{93}). The physical containts does not
change because the auxiliary fields just extend the phase space and
allocate themselvs in the unphysical sector of the theory.

The effective action corresponding to the Hamiltonian(\ref{94}) is
\begin{equation}
S_{eff} = \int dx[\pi_0\dot A_0 + \pi_1\dot A_1 +\pi_\phi\dot \phi
+ \pi_\theta \dot\theta -{\tilde {\cal H}}].\label{95}
\end{equation}
If the momenta are redefined by
\begin{equation}
\pi_1 = \dot A_1 - A'_0,
\end{equation}
\begin{equation}
\pi_0 = e\theta,
\end{equation}
\begin{equation}
\pi_\phi = \dot\phi - eA_1,
\end{equation}
\begin{equation}
\pi_\theta = \dot\theta - \theta'.
\end{equation}

It can be shown that the effective action (\ref{95}) reduces to

\begin{eqnarray}
S_{eff} &=& \int dx[{1\over2}(\partial_\mu\phi)(\partial^\mu\phi) +
e\epsilon_{\mu\nu}\partial^\mu\phi A^\nu - {1\over 4}F_{\mu\nu}F^{\mu\nu}
-{1\over 2}ae^2A_\mu A^\mu\nonumber \\
&-&{1\over 2}a((\partial_\mu\eta)(\partial^\mu\eta)) + e\partial_\mu\eta A^\mu
,\end{eqnarray}
where $\eta= a\theta$.

\section{Gauge Invariant Reformulation without \hfil \break
extending the phase space}

\par The idea was first developed by Mitra and Rajaraman(\cite{mr1,mr2}. The
method
based
on the constrainet structure of the theory. Depending on the constrint
structure
one can get different gauge invariant  action. Obviously, the physical
containts
are same in all the geuge invariant reformulants. Their methods are
applicable
in both the single chain and multi chain constrained systems. The main
idea is
to reduce the half of the constraints from a second class constrained system
keeping the first class constraints only. This new generalized Schwinger model
contains only two second class constraints and there is only one possiblity
to make it gauge invariant which is to eleminate the constraint $\omega_2$
keeping $\omega_1$ as usual i.e., to change the Hamiltonian in such a way
that $\dot\omega_1 = 0$ is satisfied. The Hamiltonian  will be of the form
\begin{equation}
{\tilde{\cal H}} = {\cal H} + {d\over e^2}(\pi'_1 -e\phi' -e^2A_0)^2.
\end{equation}
The codition $\dot\omega_1 = 0$ requires $d={1\over a}$.

The new equations of motion corresponding to the new Hamiltonian
${\tilde{\cal H}}$ are
\begin{equation}
\dot\phi = \pi_\phi - eA_1,
\end{equation}
\begin{equation}
\dot A_0 = v_0,
\end{equation}
\begin{equation}
\dot A_1 = \pi_1 - A'_0 - {1\over{e^2a}}(\pi'_1 - e\phi' - ae^2A_0)'.
\end{equation}

The first order Lagrangian is

\begin{eqnarray}
\tilde{\cal L} &=& {1\over2}(\dot\phi^2 - \phi'^2) +
e(A_1\dot\phi - A_0\phi')
-{1\over 2}ae^2(A_0^2 - A_1^2) \nonumber \\ &+& {1\over a}(\phi' + eA_0)^2
+ {1\over 2}\pi_1(\dot A_1 + {1\over {ae}}\phi'')
\label{100}\end{eqnarray}

The gauge transformation generator is $\lambda\int dx \pi_0$, where $\lambda$
is a c-number which trans form $A_0$ as $-\lambda$ and the other fields
remain unchanged. It can be shown easily that the total change in the
Lagrangian due to the above transformation $\Delta{\tilde{\cal L}} = 0$.
So the Lagrangian (\ref{100}) is gauge indariant.

\section{BRST Invariant Reformulation}

{}~~~~The BRST tecnique is to enlarge the Hilbart space of a gauge theory and
to
replace the notation of gauge transformation which shifts the operator by
C-number function with BRST transformation which mixes operator having
different statistics. It is very effective when one tries to discuss the
renormalization property of a theory. One generaly exploit the BRST symmetry
instead of exploiting the original gauge symmetry.

The combined formalism of Batalin ,Fradkin and Vilkovisky \cite{bf,bv,fv} for
quantization of a system is based on the idea that a system with second class
constraint can be made effectively first class in the extended phase space.
The field needed for this conversion turns out to the Wess-Zumino scalar with
the proper choise of gauge condition, as pointed out by FIK.

If a canonical Hamiltonian in terms of the canonical pairs
$(p_i, q^i),  i=1,2.......N$ is considered subjected to  a set of constraints
$\Omega_a = 0,  a=1,2.......n$ and it is assumed that the constraints satisfy
the following algebra.

\begin{equation}
[\Omega_a, \Omega_b] = i\omega_cU^c_{ab},
\end{equation}
\begin{equation}
[H_c, \Omega_a] = i\Omega_bV^b_c,
\end{equation}

then m no of additional condition $\Phi_a = 0$ with
 $det[\Phi_a, \Omega_b] \neq 0$ have to be imposed inorder to single out the
physical degrees of freedom.

The Hamiltonian consistent with the set of constraints $\Omega_a = 0$ and
$\Phi_a = 0$ together with Hamiltonian equation of motion  is obtained from
the action

\begin{equation}
S = \int dt[p_iq^i - H_c(p_i,q^i) - \lambda^a\Omega_a + \pi_a\Phi^a],
\end{equation}

where $\lambda^a$ and $\pi_a$ are lagrrangian multiplier field satisfying the
relation $[\lambda^a, \pi_b] = i\delta^a_b$.

Now introducing one pair of canonical ghost field $(C^a,\bar{P}_a)$ and one
pair of canonical antighost field $(P^a,\bar{C}_a)$ for each pair of
constraints
an equivalance can be made to the initial theory of the unconstrained phase
space. So the quantum theory can be described by the action
\begin{equation}
S_{qf} = \int dt[p_iq^i + \pi^a\dot\lambda_a + \bar{P}^a\dot C_a +
+ \bar{C}^a \dot P_a- H_{BRST} + i[Q, \psi].
\end{equation}

$H_{BRST}$ is the minimal Hamiltonian as termed by Batalin and fradkin,
given by
\begin{equation}
H_{BRST} = H_c + \bar{P}_aV^a_bC^b.
\end{equation}

The BRST charge and the fermionic gauge fixing function are given by
\begin{equation}
Q = C^a\omega_a - {1\over 2}C^bC_cU^c_{ab} - P^a\pi_a,
\end{equation}
\begin{equation}
\psi = \bar{C}_c\chi^a + \bar{P}^a\lambda^a,
\end{equation}
where $\chi_a$ are given by the gauge fixing condition $\Phi_a =\dot\lambda_a
 + \chi_a$.

The model consider here is described by the Lagrangian (\ref{3}), The
canonical Hamiltonian and the momenta are given in (\ref{4}),(\ref{3}),
(\ref{31}) and(\ref{32}). The two second class constraints are
\begin{equation}
\omega_1 = \pi_0,
\end{equation}
\begin{equation}
\omega_2 = \pi_1' - e\phi'- ae^2A_0.
\end{equation}
Now introducing BF field $\theta$ and $\pi_\theta$ the constraints are made
first class
\begin{equation}
\tilde\omega_1 = \omega_1 - e\theta,\label{16}
\end{equation}

\begin{equation}
\tilde\omega_2 = \omega_2 + ea\pi_\theta.\label{17}
\end{equation}
The time involution of the constraints are
\begin{equation}
\dot\omega_1 = i\omega_2, \label{18}
\end{equation}
\begin{equation}
\dot\omega_2 = i\omega''.\label{19}
\end{equation}

In general the first class Hamiltonian consistent consistent with the
constraints ${\tilde\omega_1}$ and ${\tilde\omega_2}$ will be the original
Hamiltonian added with a polynomial of BF field. And the polynomial will
be determined by the condition that the new first class constraints will
satisfy the same time involution like the old second class set of constraint
(\ref{18}) and (\ref{19})The extra term $H_{BF}$ is found out to be

\begin{equation}
H_{BF} = -{1\over 2}(\pi^2_\theta + {1\over a^2}\theta'^2).
\end{equation}

The BRST charge and the fermionic gauge fixing condition are

\begin{equation}
Q = \int dx [B_1P^1 + B_2P^2 + C_1{\tilde\omega}^1 + C_2{\tilde\omega}^2],
\end{equation}
\begin{equation}
\psi = \int dx[\bar{C}_1\chi^1 + \bar{C}_2\chi^2 + \bar{P}_1N^1
+ \bar{P}_2N^2].
\end{equation}

Here the gauge fixing function are chosen to be $\chi_1 = A_0$ and $\chi_2 =
\partial^1A_1 + {\alpha\over 2}B_2$.
It is easy to check that

\begin{equation}
[Q, H_{BRST}] = 0
\end{equation}
\begin{equation}
[[\psi, Q], Q] = 0
\end{equation}
\begin{equation}
Q^2 = [Q, Q] = 0
\end{equation}

The effective action is given by
\begin{eqnarray}
S_{eff} &=& \int dx[\pi^0\dot A_0 + \pi^1\dot A_1 + \pi_\phi\dot\phi +
\pi_theta\dot\theta + B_2\dot N^2  \nonumber \\
&+&\bar{P}_1\dot C^1 + \bar{P}_2\dot C^2
+ \bar{C}_2\dot P^2 - H_{tot},
\end{eqnarray}
where $H_{tot} = H_{BRST} - i[Q, \psi]$.

Here the term $\int dx[B_1\dot N^1 +
\bar{C}_1\dot P^1 = i[Q, \int dxB_1\dot N^1]$ is suppressed in the Legender
transformation by replacing $\chi_1$ with $\chi_1 + \dot N_1$. The generating
function is now given by
\begin{equation}
Z= \int[DM]exp(iS_{eff}),
\end{equation}
 where
\begin{eqnarray}
[DM]&=&[DA_0 D\pi^0][DA_1 D\pi^1][D\phi D\pi_\phi][D\theta D\pi_\theta]
\nonumber \\ &\times&[DC^1 D\bar{P}_1][DC^2 D\bar{P}_2][DP^1 D\bar{C}_1]
\nonumber \\ &\times&[DP^2 D\bar{C}_2][DN^1 DB_1][DN^2DB_2]
\end{eqnarray}

In order to derive the covariant action one should eliminate $N_1$, $B_1$,
$C^1$, $\bar{C}_1$,  $P^1$, $\bar{P}_1$, $A_0$ and $\pi_0$ by Gaussian
integration.After integration the effection reduces to

\begin{eqnarray}
S_{eff} &=& \int dx[\pi_0\dot A_0 + \pi_1\dot A_1 +\pi_\phi\dot \phi
+ \pi_\theta \dot\theta + B\dot N + \bar{P}\dot C + \bar{C}\dot P \nonumber \\
&-& [{1\over 2}(\pi^2_1 + \pi^2_\phi + \phi'^2 - eA_1\pi_\phi -
{1\over 2}(a - 1)e^2A^2_1] - ea\theta'\nonumber \\
&+&{a\over2}(\pi^2_\theta + {1\over a^2}
\theta'^2)  + B(A'_1 + {a\over 2}B)\nonumber \\ &+& N(\pi'_1 - e\phi'
+ ea\pi_\theta)-C\bar{C}''.
\end{eqnarray}
Again integrating over $\pi_1, \pi_\phi, \pi_\theta$ and $\bar{P}$ one can
obtain
\begin{eqnarray}
S_{eff} &=& \int dx[{1\over2}(\partial_\mu\phi)(\partial^\mu\phi) +
e\epsilon_{\mu\nu}\partial^\mu\phi A^\nu - {1\over 4}F_{\mu\nu}F^{\mu\nu}
-{1\over 2}ae^2A_\mu A^\mu \nonumber\\
&-&{1\over 2}a((\partial_\mu\eta)(\partial^\mu\eta)) + e\partial_\mu\eta A^\mu
\nonumber \\
&-&\partial_\mu\bar{C}\partial^\mu C + A_\mu\partial^\mu B - {\alpha\over
2}B^2.
\end{eqnarray}
where $N$ is replaced by $-A_0$ and $\eta$ is defined by $\eta = a\theta$.
The action is invariant under the BRST transformation $\delta X = \lambda\int
dx [Q, X]$ where $X$ stands for the field variables of the above action and
$\lambda$ is a grassmanian parameter.

\section{Conclusion}

{}~~~~To summarize,  we have looked at the familiar Schwinger model in
the one parameter class of regularizations used in
studies on  the  chiral
Schwinger  model  and shown that  although only global gauge invariance
is maintained, the theory is sensible in  every  way  and  can
be  solved  exactly.  The  spectrum  consists  of the usual
massive boson with  the  mass  explicitly  depending  on  the
regularization  parameter  --  but  there  is  also a massless
fermion, which is the main difference from the usual treatment.

Gauge invariant reformulation is done in two ways one in the extended phase
spaces and the other in the original phase space both the cases the gauge
invariant action is superficially different from the actual action but the
physical containts are alike. The external field introduced to enlarge the
phase space is shown to be identical to the Wess-Zumino scalar when
Woutzasec's formalism is considered. But in case of Mitra and Rajaraman' s
formalism a gauge invariant action is obtained without the emergence of any
Wess-Zumino term.

Following BVF formalism a BRST invariant action is obtained. In the usual
Hamiltonian formalism of a gauge invariant theory one some times need to
destroy the gauge symmetry under the introduction of some gauge fixing terms.
However, BRST invariant Hamiltonian which has beeen reformulated will help
one to work in an extended phase space on which only a subspace corresponds to
the state of physical interest.

The Schwinger model is well known to be solvable, the new generalization
differs from the usual one by the mass terms of the gauge fields do not
loose the solvablity. A gauge invariant as well as a BRST invariant
reformulation is made so that calculations ordinarily made with the gauge
invariant Schwinger model can be repeted here. New physical consequences
will arise, as is indicated by the fact that even the spectrum gets altered.
\vfil\eject

\end{document}